\documentclass[%
 preprint,
superscriptaddress,
 amsmath,amssymb,
 aps,
pra,
]{revtex4-1}

\usepackage{graphicx}% Include figure files
\usepackage{dcolumn}% Align table columns on decimal point
\usepackage{bm}% bold math
\usepackage{hyperref}% add hypertext capabilities
\begin{document}

%\preprint{APS/123-QED}

\title{Cavity ring-up spectroscopy for  dissipative and dispersive sensing in a whispering gallery mode resonator}% Force line breaks with \\

\author{Yong Yang}
   \email{yong.yang@oist.jp}
 \affiliation{Light-Matter Interactions Unit, Okinawa Institute of Science and Technology Graduate University, Onna, Okinawa 904-0495, Japan}
  \affiliation{National Engineering Laboratory for Fiber Optics Sensing Technology, Wuhan University of Technology, Wuhan, 430070, China
}%

\author{Ramgopal Madugani}%
\affiliation{Light-Matter Interactions Unit, Okinawa Institute of Science and Technology Graduate University, Onna, Okinawa 904-0495, Japan}
\affiliation{Physics Department, University College Cork, Ireland}


\author{Sho Kasumie}
 \affiliation{Light-Matter Interactions Unit, Okinawa Institute of Science and Technology Graduate University, Onna, Okinawa 904-0495, Japan}
\author{Jonathan M. Ward}
\affiliation{Light-Matter Interactions Unit, Okinawa Institute of Science and Technology Graduate University, Onna, Okinawa 904-0495, Japan}
\author{S\'ile Nic Chormaic}
\affiliation{Light-Matter Interactions Unit, Okinawa Institute of Science and Technology Graduate University, Onna, Okinawa 904-0495, Japan}

\date{\today}% It is always \today, today,

\begin{abstract}
In whispering gallery mode resonator sensing applications, the conventional way to detect a change in the parameter to be measured is by observing the steady state transmission spectrum through the coupling waveguide. Alternatively, cavity ring-up spectroscopy (CRUS) sensing can be achieved transiently. In this work, we  investigate  CRUS using coupled mode equations and find analytical solutions with a large spectral broadening approximation of the input pulse. The relationships between the frequency detuning, coupling gap and ring-up peak height are determined and experimentally verified using an ultrahigh \textit{Q}-factor silica microsphere. This work shows that distinctive dispersive and dissipative transient sensing can be realised by simply measuring the peak height of the CRUS signal, which might improve the data collection rate.
\end{abstract}

\pacs{Valid PACS appear here}% PACS, the Physics and Astronomy
\maketitle

%\tableofcontents
\section{Introduction}
\label{sec:intro}
Whispering gallery mode (WGM) resonators are widely used for a number of applications, one of which is sensing \cite{Ward2014,Foreman2015}. The high optical quality factor (\textit{Q}-factor) and relatively small mode volume of  whispering gallery resonators (WGRs) renders the modes very sensitive to subtle environmental changes. Until now, WGRs have been used to measure changes in a number of parameters such as refractive index \cite{Hanumegowda2005,White2006}, temperature \cite{Dong2009,Yan2011,Ward2013}, pressure \cite{Henze2011,Yang2016} and stress\cite{Sumetsky2010,Madugani2012}. Aside from parameter change deteection, ultrahigh \textit{Q} resonators have also been used to detect nanoparticles \cite{Li2014,Ozdemir2014} and single viruses \cite{Vollmer2008,He2011}. The mechanism behind ultrahigh sensitivity sensing in WGRs is based on a reactive (i.e. dispersive) frequency shift of the whispering gallery modes \cite{Vollmer2008} as a result of perturbations that may be present. Alternatively,  a perturbation may increase the optical linewidth of the WGM by introducing more dissipation \cite{Hu2014}, or may change the observed mode splitting if modal coupling is present \cite{Li2014,He2011,Knittel2013}. The optomechanical properties of  WGRs can also be used for force \cite{Gavartin2012} or viscosity sensing \cite{Bahl2013}.

Currently, in order to retrieve the dispersive, dissipative and mode splitting information, the transmission spectrum of a WGR through an externally-coupled waveguide, such as a tapered optical fibre, is usually measured. Light from a tunable laser source is coupled into the tapered fibre and the transmission is monitored.  Low powers are used in order to minimise thermal and nonlinear effects on the whispering gallery modes.  By sweeping the laser frequency, the transmission spectrum through the fibre can be recorded. Any  changes to the frequency, mode splitting, or linewidth are used to monitor  perturbations induced by the physical parameter that is being sensed. During measurements, the transmission spectrum represents a \textit{steady state} of the coupled system due to limitations on the scanning speed, thereby constraining the time response of the sensor \cite{Stern2012,Rosenblum2015,Shu2015}. For a WGR with an  optical \textit{Q}-factor $>2\times10^7$, a ringing effect is observable even if the laser is scanned as quickly as 100 Hz \cite{Dong2009ringing}. The ringing spectrum can be used to distinguish between the over-coupling and under-coupling cases \cite{Rasoloniaina2014}. When the scanning speed is lower than the \textit{character speed}, as defined in \cite{Shu2015}, the steady state treatment can no longer be used to describe the coupled mode system. Therefore, by recording lineshape changes in the ringing tail of an observed transmission spectrum, either by (i) a scanning probe laser or (ii) a fixed laser in resonance with a high \textit{Q} mode, transient sensing should be possible \cite{Shu2015}. A proof-of-principle experiment based on the ringing phenomenon has recently been reported \cite{Ye2016}.

Another possible approach is to send light pulses, which are far detuned from a WGM resonance, through the optical coupler. The retrieved signal on the coupler's output  shows an oscillatory lineshape similar to that in \cite{Shu2015}; this effect is termed \textit{cavity ring-up spectroscopy} (CRUS) \cite{Rosenblum2015} and the rising edge of the light pulse leads to transient broadening. Even though the light is far detuned from the WGM, a fraction can still be coupled into the cavity if the broadening is much larger than the detuning.  The system is not affected by thermal or nonlinear processes which may arise due to the ultrahigh \textit{Q} of the mode. The ringing effect occurs within the lifetime of the WGM and, therefore, can be used for ultrafast sensing. The transient capability of CRUS has already been demonstrated by measuring the time response for thermo-refractive effects, Kerr nonlinearity and optomechanical vibrations \cite{Rosenblum2015}.

To date, there has been no thorough theoretical investigation of CRUS and details, such as the influence of the pulse's rise time on the observed spectra, are relatively unknown. In this manuscript, we use coupled mode theory to solve the related differential equations without relying on the steady state assumption. An approximate analytical solution is obtained and compared to a precise numerical transient solution. The theoretical results fit well to experimental data that we obtained for a silica microsphere resonator. The influences of the pulse rise time, the coupling condition and the detuning on the ring-up spectrum are given.  This provides a solid foundation for future applications in transient sensing.

\section{Coupled mode theory}
\label{sec:theory}
\begin{figure}
\resizebox{0.75\textwidth}{!}{\includegraphics{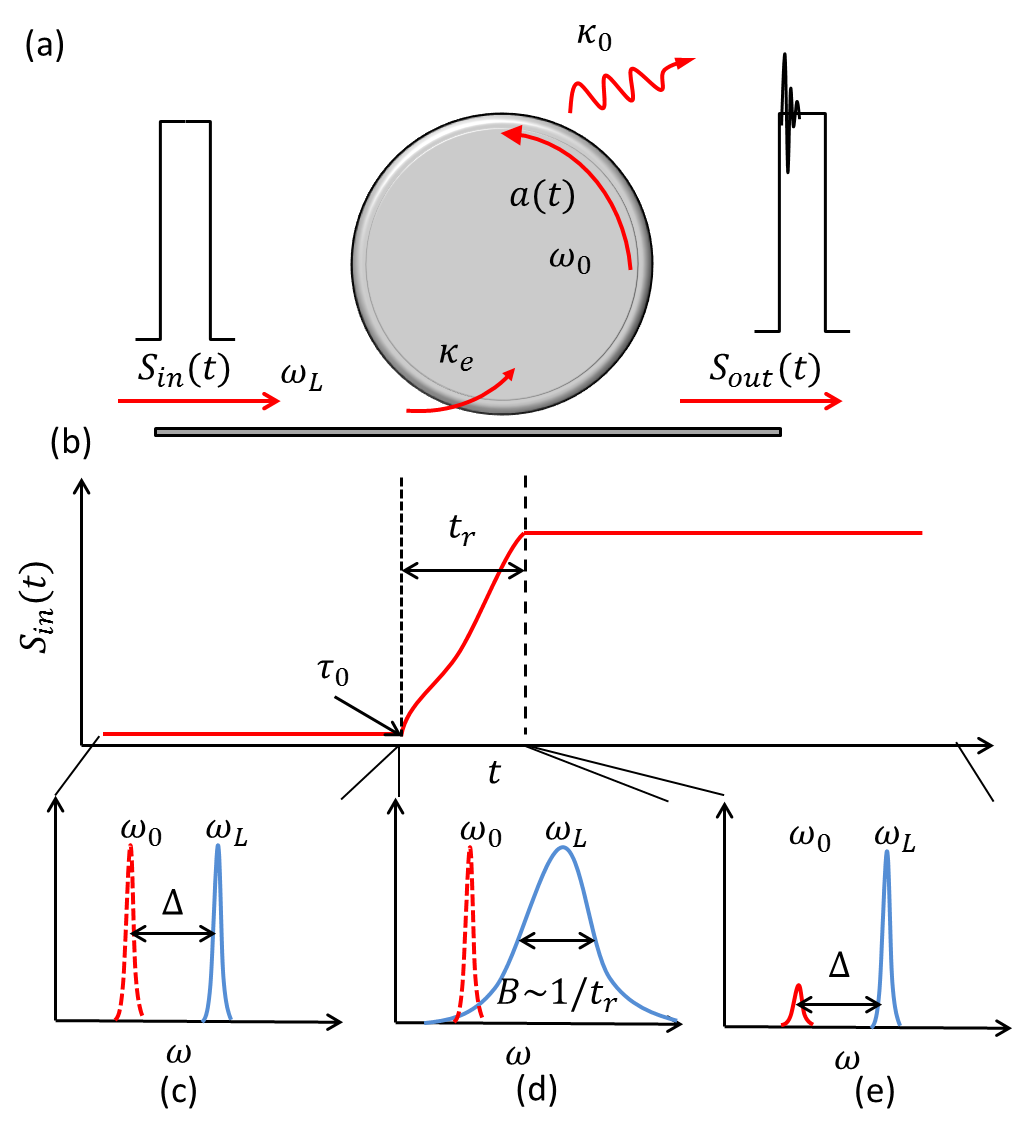}}
\caption{(a) Schematic of a taper coupled WGR system for transient sensing using CRUS. A pump laser of frequency $\omega_L$, far off resonance with the WGM, $\omega_0$, is coupled through the tapered fibre with a temporal profile, $S_{in}(t)$. The pulse profile is depicted in (b); the laser pulse starts at a time $\tau_0$ and rises up to its maximum within a time $t_r$. The mathematical description of the pulse is given in Eq. \ref{eq:pulse}. (c)-(e) Transient frequency of the laser pulse for different time intervals.  (c): $t=[0, \tau_0]$; (d): $t=[\tau_0,\tau_0+t_r]$; (e): $t=[\tau_0+t_r,+\infty$). At the rising edge of the pulse, the laser source is transiently broadened (d), so a fraction of the pump signal couples to the WGM and contributes to a beat signal between it and the pump source (e).}
\label{fig:scheme}
\end{figure}

A typical CRUS setup is shown in Fig. \ref{fig:scheme}(a). The WGR is coupled evanescently to a tapered optical fibre through which light from a laser propagates.  The light couples into the resonator and is monitored at the opposite end of the fibre.   The coupling dynamics can be described using coupled mode theory. The amplitude of the intracavity electromagnetic field, $a(t)$, changes in time according to the following \cite{Shu2015,Gorodetsky1999,Zou2008}
\begin{equation}
\frac{da(t)}{dt}=-j\omega_0a(t)-(\kappa_e+\kappa_0)a(t)+\sqrt{2\kappa_e}S_{in}(t),
\label{eq:couplemode}
\end{equation}
where $j = \sqrt{-1}$, the resonant frequency of the WGM is $\omega_0$, and $\kappa_e$ and $\kappa_0$ represent the external and intrinsic coupling rates, respectively. The total damping rate of the cavity is given by $\kappa=\kappa_e+\kappa_0$ and $\tau=1/\kappa$ is the intracavity lifetime. In order to arrive at the transient response of the WGR, the laser light is pulsed with a temporal profile, $S_{in}(t)$.  The laser frequency, $\omega_L$, is far detuned, i.e. $\omega_L-\omega_0\gg\kappa$. The pulsed input field can be separated into a slowly varying and a fast varying term so that
\begin{equation}
S_{in}(t)=a_{in}(t)e^{-j\omega_Lt}.
\label{eq:input}
\end{equation}
\noindent Here, the slowly varying part, $a_{in}(t)$, represents the temporal profile of the pulse and, in the following discussions, it takes the form of a Guassian function where
\begin{equation}
a_{in}(t)=\left\{
\begin{aligned}&0,& t<\tau_0 \\ &\alpha_{in}exp\left(-\frac{4[t-\tau_0-t_r]^2}{ln2\cdot t_r^2}\right) , & \tau_0<t<\tau_0+t_r \\ &\alpha_{in}, & t>\tau_0+t_r
\end{aligned}\right.
\label{eq:pulse}
\end{equation}
The pulse is illustrated in Fig. \ref{fig:scheme}(b). The pulse starts at time $\tau_0$ and follows a Gaussian profile with a rise time, $t_r\ll\tau$. At time $t=\tau_0+t_r$, the total power of the pulse reaches its maximum, $|\alpha_{in}|^2$, and, for later times, the laser can be treated as a continuous light source over the lifetime of the cavity mode.

\subsection{A simple model: large spectral broadening bandwidth}
\label{sec:simple}
The temporal profile of $S_{in}(t)$ can be obtained from Eq. \ref{eq:input} and Eq. \ref{eq:pulse} and Fourier expanded as follows:
\begin{equation}
S_{in}(t)=\left\{
\begin{aligned}&0, & t<\tau_0\\ &\sqrt{\frac{ln2}{\pi}}\frac{t_r\alpha_{in}}{4}\int_{-\infty}^{+\infty}e^{-\frac{ln2 t_r^2\omega^2}{16}}e^{-j[(\omega+\omega_L)(t-\tau_0)]}d\omega, & \tau_0<t<\tau_0+t_r\\ &\alpha_{in}e^{-j\omega_Lt}, & t>\tau_0+t_r
\end{aligned}\right.
\label{eq:fourier}
\end{equation}
It can be seen that, for a time interval $t\in[\tau_0, \tau_0+t_r]$, the rise time of the pulse induces sideband frequencies even though the laser source is monochromatic. As the Fourier transform of a Gaussian function is a Gaussian, the laser pulse is expanded transiently with a bandwidth, $B=1/(t_rln2)$.

\begin{figure}
\resizebox{0.75\textwidth}{!}{\includegraphics{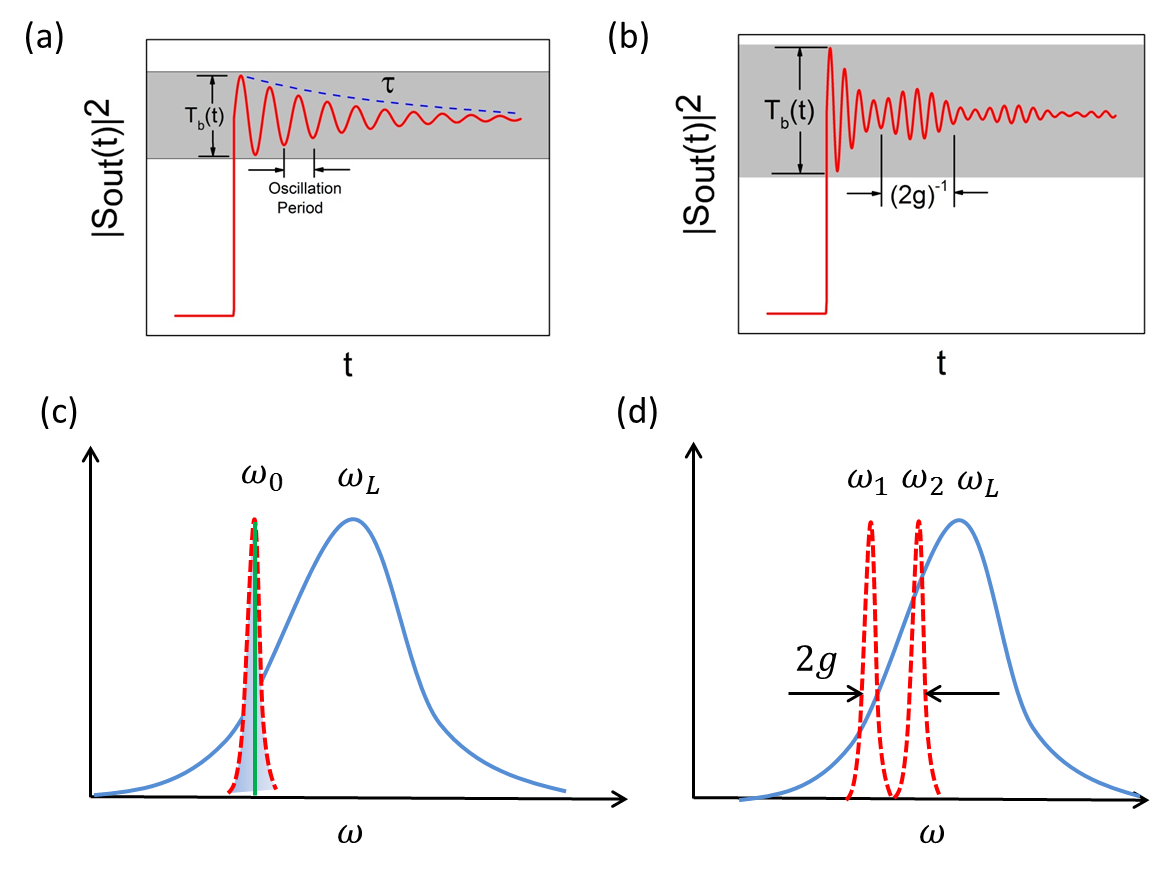}}
\caption{(a) A typical CRUS signal. The curve in the gray rectangle is the beat signal deduced from Eq. \ref{eq:beat}. This is due to the frequency broadening at the rising edge of the input signal. We can treat the WGM as an ultra-narrow filter; therefore, only the on-resonance, $\omega=\omega_0$, component in the broadened pump laser signal can be stored in the WGM as shown in (c). (b) The CRUS signal when there is modal coupling with  coupling strength, $g$. The beating is modulated by a cosine envelope of period $(2g)^{-1}$. This is due to the beating between the two normal standing wave modes, with resonant frequencies of $\omega_1$ and $\omega_2$, as shown in (d).}
\label{fig:CRUS}
\end{figure}

In our experiments, the pulse has a rise time ranging from  $\sim$1 ns to several tens of ps. The laser frequency broadening bandwidth, $B$, is of the order of GHz, and, for a WGM with $Q>10^7$, $\kappa\sim$ MHz. As $B\gg\kappa$, we can assume that only the portion of the broadened laser source at the resonant frequency, $\omega_0$, can be efficiently coupled to the WGM. As illustrated in  Fig. \ref{fig:scheme}(c)-(e), we assume that the laser is far red-detuned from the resonant frequency, such that $\omega_L-\omega_0=\Delta\gg\kappa$. The WGM acts as an infinitely narrow, band pass filter so that we only need to consider the frequency at $\omega_0$, see Fig. \ref{fig:CRUS} (c). Thus the input signal can be simplified as:
\begin{equation}
S_{in}(t)\approx \sqrt{\frac{ln2}{\pi}}\frac{t_r\alpha_{in}}{4}e^{-\frac{ln2t_r^2\Delta^2}{16}}e^{-j\omega_0t}.
\label{eq:simple}
\end{equation}
Note that, in the above expression, only the last factor is time dependent.  The single frequency pulse as defined by Eq. \ref{eq:simple} must satisfy Eq. \ref{eq:couplemode}.  Thence, we get the dynamics of the intracavity amplitude for the transient interval at the rise time and for later times. Eq. \ref{eq:couplemode} has a general solution \cite{Shu2015}:
\begin{equation}
a(t)=\sqrt{2\kappa_e}a_{in}e^{j\omega_0t-\kappa t}[\frac{\tau}{1+j(\omega_L-\omega_0)\tau}+ \int_{\tau_0}^{t}e^{j\phi(t')-j\omega_0t'+\kappa t'}dt'].
\label{eq:general}
\end{equation}

The complete dynamics can be separated into two steps. First, at the rise time interval, the signal is broadened and light of frequency  $\omega_0$ is coupled into the cavity. As $t_r\ll\tau$,  dissipation during this short time can be ignored and $\omega_L$ in Eq. \ref{eq:general} can be substituted by $\omega_0$. If we consider the accumulated phase  $\phi(t')=\int_{\tau_0}^{t'}\omega_0(t")dt"=\omega_0t'$, then the accumulated amplitude at a time $t=\tau_0+t_r$ is given by
\begin{equation}
a(t)=\frac{\sqrt{2\kappa_e}}{\kappa}\sqrt{\frac{ln2}{\pi}}\frac{t_r\alpha_{in}}{4}e^{-\frac{ln2t_r^2\Delta^2}{16}}e^{-j\omega_0t}, \tau_0<t<t_r.
\label{eq:riseup}
\end{equation}

Next, as already explained,  the frequency returns to $\omega_L$ and is far detuned. Therefore, for later times, there is no light coupled into the WGM and the system follows simple decay dynamics with a decay rate, $\kappa$, whose initial value is as in Eq. \ref{eq:riseup}:
\begin{equation}
a(t)= \frac{\sqrt{2\kappa_e}}{\kappa}\sqrt{\frac{ln2}{\pi}}\frac{t_r\alpha_{in}}{4}e^{-\frac{ln2t_r^2\Delta^2}{16}}e^{-j\omega_0t-\kappa t}, t>\tau_0+t_r.
\label{eq:decay}
\end{equation}
The output signal, $S_{out}(t)$, can be calculated using the input-output relationship:
\begin{equation}
S_{out}(t)=S_{in}(t)-\sqrt{2\kappa_e}a(t).
\label{eq:inout}
\end{equation}

Here, we are interested in the output signal after the pulse reaches its maximum value. At this point the input signal changes to a continuous state and the broadening vanishes, yielding $S_{in}(t)=\alpha_{in}e^{-j\omega_Lt}$. It can be seen from Eq. \ref{eq:decay} and Eq. \ref{eq:inout} that the final signal is a superposition of the laser signal at $\omega_0$, which is the residue of intracavity photons leaking out after the rise time, and $\omega_L$ of the pumping signal. Therefore, it gives transient transmission, $T(t)$, at a time after the rise up of the pulse as follows:
\begin{equation}
T(t)=|\frac{S_{out}}{S_{in}}|^2=1+\frac{\kappa_e}{\kappa}\sqrt{\frac{ln2}{\pi}}e^{-\frac{ln2t_r^2\Delta^2}{16}}e^{-\kappa t}\sin{(\Delta (t-\tau_0-t_r))}.
\label{eq:beat}
\end{equation}

A typical waveform is shown in Fig.\ref{fig:CRUS}(a). There are three values that can be retrieved from the CRUS signal: peak height, oscillation period, and the decay rate. From Eq. \ref{eq:beat}, the oscillation frequency is the detuning frequency, $\Delta$ and the decay rate is the lifetime, $\tau$.  Note that this is the loaded\textit{ Q}-factor of the system. The peak height is more complicated as it is related to both the dissipation rate and the detuning.   This will be discussed in more detail later.

\subsection{Transient response during the rising edge}
\label{sec:fardetune}
From the above discussion, we assume that the phase accumulation during the rise time can be ignored.  This is not valid when considering a large value for the detuning. As shown in Eq. \ref{eq:beat}, the frequency of the obtained transient signal is determined by the detuning of the laser source from the resonant frequency. At large detuning, one period of oscillation is almost as long as the rise time; hence, the phase accumulated during the rise time can influence the output signal.  This distorts the first few periods of the beat signal. To calculate this, we should use the full form of Eq. \ref{eq:riseup} at the rise time.

The output signal, $S_{out}(t)$, can be determined from Eq. \ref{eq:inout}, with the input amplitude defined in Eq. \ref{eq:pulse}. Therefore, a beating frequency equaling $\Delta=\omega_L-\omega_0$ can be achieved during the rise time :
\begin{equation}
T(t)\approx[1+\sqrt{\frac{ln2}{\pi}}\frac{\kappa_e}{2\kappa}e^{-\kappa(t-\tau_0)}\sin{(\Delta (t-\tau_0))}]e^{-\frac{4(t-\tau_0-t_r)^2}{ln2\cdot t_r^2}}.
\label{eq:beat_at_riseup}
\end{equation}

The signal which arrives at the detector can be separated into two components. The first part has the same form as that in Eq. \ref{eq:pulse}, i.e. the rising part of the retrieved signal maintains a Gaussian profile and the rise  time is unchanged. The second part comes from the contribution of the intracavity amplitude and takes the form of a trigonometric function with a period  $2\pi/\Delta$. Since this component only occurs for a time interval, $t_r$, if $\Delta$ is small it follows that $\sin{(\Delta(t-\tau_0))}\approx 0$ and the rising edge waveform will not be disturbed.

\subsection{The mode mixing case}
\label{sec:mix}
In practice, multiple WGM resonances may fall within the transient spectral broadening bandwidth. In the following, the case in which two modes mix  will be investigated. There are two different types of mode mixing that we consider.  The first situation is that where two modes exist quite close to each other in the spectrum. In the following discussion, we assume that  both modes lie within the transiently broadened spectrum bandwidth, $B$, while satisfying the narrow band pass filter approximation made in Sec. \ref{sec:simple}. Defining the two modes as $\omega_1$ and $\omega_2$, with a separation between them of $\delta$, the coupled mode equations for each mode are given by
\begin{equation}
\frac{da_i}{dt}=-j\omega_ia_i(t)-(\kappa_{0,i}+\kappa_{e,i})a_i(t)+\sqrt{2\kappa_e}a_{in,i}(t)e^{-j\omega_it},
\label{eq:couple2}
\end{equation}
where $i=1,2$ and represents the two modes. Since there is no spectral overlap of the two modes, the mode equations are decoupled and can be solved separately. This yields the same expressions as in Eq. \ref{eq:decay} and each mode beats with $S_{in}(t)$ separately.   The total beat signal, $T_b(t)$, is
\begin{equation}
T_b(t)=\sum_{i=1,2}T_{i,b}(t)=\sum_{i=1,2}\frac{\kappa_{e,i}}{\kappa_i}\sqrt{\frac{ln2}{\pi}}e^{-\frac{ln2t_r^2\Delta_i^2}{16}}e^{-\kappa_it}\sin{(\Delta_it)}.
\label{eq:uncoupled}
\end{equation}
$T_b(t)$ is a linear combination of the two individual beat signals. The amplitudes of the signals with different frequencies are determined from the detunings, $\Delta_i$, of the resonances relative to the laser source. If $\delta$ is large, then one of the beat signals will be much larger than the other since $T_{b,i}$ follows the Gaussian relationship with the detuning. In this case, the weaker signal can be ignored. In contrast, if the two modes are close to each other so that $\delta\ll\Delta_i$, then, for simplicity, we set $\Delta_1\approx\Delta_2=\Delta$ and the coupling rates are approximately equal. From Eq. \ref{eq:uncoupled}, the total beat signal, $T_b(t)$ ,is ($T(t)=1+T_b(t)$):
\begin{equation}
T_b(t)\approx\frac{\kappa_e}{\kappa}\sqrt{\frac{ln2}{\pi}}e^{-\frac{ln2t_r^2\Delta^2}{16}}e^{-\kappa t}(1+\cos{(\delta t)})\sin{(\Delta t)}.
\label{eq:uncoupledbeat}
\end{equation}

The other possible case to consider is that where the two modes are coupled. This is often the case in traveling wave resonators, such as WGRs.  Degenerate modes in the resonator represent  clockwise and counter clockwise propagation. Due to  scattering along the propagation path, the two modes can indirectly couple to each other; this effect is called \textit{modal coupling} and leads to normal mode splitting \cite{Kippenberg2002,Srinivasan2007}. Here, we define a modal coupling strength, $g$, and the  coupled mode equations for the two modes are given by $a_{cw}$ and $a_{ccw}$ representing the two opposing propagation directions. The two coupled mode equations can be written as
\begin{equation}
\begin{aligned}
&\frac{da_{cw}}{dt}=-j\omega_0a_{cw}(t)-(\kappa_0+\kappa_e)a_{cw}(t)-jga_{ccw}(t)+\sqrt{2\kappa_e}S_{in}(t);\\
&\frac{da_{ccw}}{dt}=-j\omega_0a_{ccw}(t)-(\kappa_0+\kappa_e)a_{ccw}(t)-jga_{cw}(t).
\end{aligned}
\label{eq:modalcouple}
\end{equation}
The two modes should have the same values of $\kappa_0, \kappa_e$ and detuning. By letting $A_1(t)=a_{cw}(t)+a_{ccw}(t)$ and $A_2(t)=a_{cw}(t)-a_{ccw}(t)$, these two equations can be transformed into two uncoupled equations \cite{Srinivasan2007}:
\begin{equation}
\begin{aligned}
&\frac{dA_1(t)}{dt}=-j(\omega_0+g)A_1(t)-(\kappa_0+\kappa_e)A_1(t)+\sqrt{2\kappa_e}S_{in}(t)\\
&\frac{dB(t)}{dt}=-j(\omega_0-g)A_2(t)-(\kappa_0+\kappa_e)A_2(t)+\sqrt{2\kappa_e}S_{in}(t)
\end{aligned}
\label{eq:sumofcme}
\end{equation}
Here, the original two travelling modes generate two equivalent standing wave modes with frequency shifts, $\pm g$, as illustrated in Fig. \ref{fig:CRUS}(d). Usually, the modal coupling strength is in the MHz range for silica WGRs, i.e. the frequencies of the two standing waves are quite close to each other. Since $B\gg 2g$, we can assume that the two modes have the same transient components at the rising edge time, as in the previous case. Also, in the actual experiments, we measure the light transmitted through the tapered fibre, so that $T(t)=|(1-\sqrt{2\kappa_e}a_{cw}(t))/S_{in}(t)|^2$, where $a_{cw}=1/2(A_1(t)+A_2(t))$. $T_b(t)$ is a beat signal between the initial signals with frequencies $\Delta$ and $\Delta\pm g$ and has the following form
\begin{equation}
T_b(t)\approx \frac{\kappa_e}{\kappa}\sqrt{\frac{ln2}{\pi}}e^{-\frac{ln2t_r^2\Delta^2}{16}}e^{-\kappa t}\sin{(\Delta t)}\{1+\frac{1}{2}\cos{(2gt)}\}.
\label{eq:modalcouplebeat}
\end{equation}
From the above equation, when there is mode splitting due to  intrinsic scattering, the transient signal has a cosine form of frequency $\Delta$ and is modulated by a slowly oscillating envelope with a frequency of $2g$. This yields a similar waveform to that obtained in the uncoupled modes' case. Comparing Eq. \ref{eq:modalcouplebeat} to Eq. \ref{eq:uncoupledbeat}, we see that there is a factor of $1/2$ difference. When modal coupling is present, the two modes interfere with each other, whereas for the uncoupled case, the beating of the two modes has no coherent property.

\subsection{Numerical method}
\label{sec:numerical}
\begin{figure}
\resizebox{0.99\textwidth}{!}{
\includegraphics{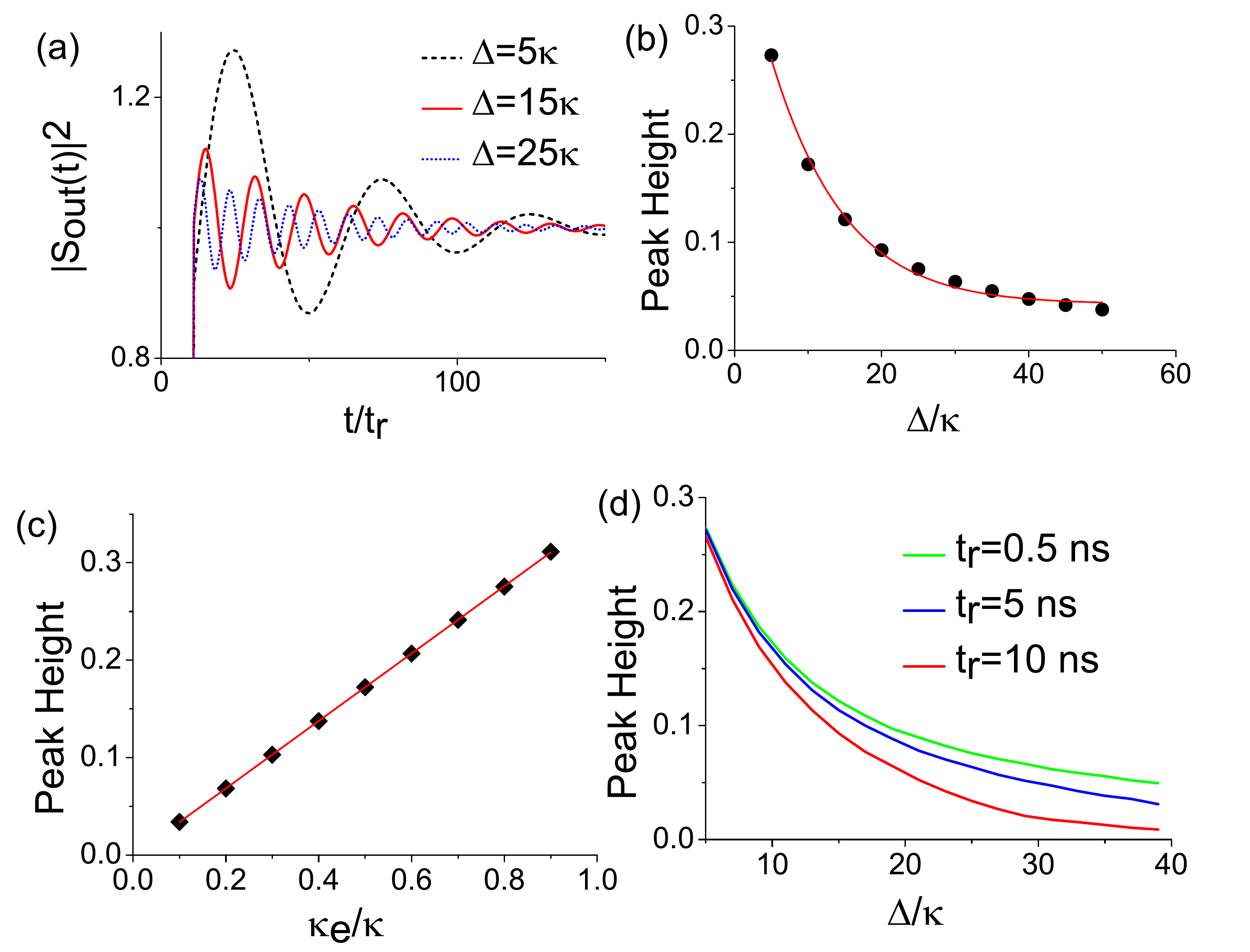}}
\caption{(a). The normalised transient response of a pulse detuned to the high Q WGM. The legend shows the detuning (normalized to $\kappa$) for different curves. (b) The peak heights of the transient signals for different detunings in (a). (c) The peak heights at different coupling conditions $\kappa_e/\kappa$. (d) Peak heights as a function of detuning for certain coupling conditions with different pulse rise-up times, $t_r$.}
\label{fig:simulate1}
\end{figure}

For a more precise simulation of the transient system, we should solve the coupled mode equations numerically. Eq. \ref{eq:couplemode} can be transformed into a rotating frame of reference, with an angular frequency $\omega_L$ such that
\begin{equation}
\frac{da(t)}{dt}=-j\Delta a(t)-\kappa a(t)+\sqrt{2\kappa_e}a_{in}(t).
\label{eq:rotateframe}
\end{equation}
The input-output relationship for the rotating frame is $a_{out}(t)=a_{in}(t)-\sqrt{2\kappa_e}a(t)$. For simplicity, we rescale the time by $t_r$ in the following discussion and we assume that the WGR is critically coupled to the external coupler, unless we explicitly mention otherwise. First, let us assume that the lifetime of a WGM is about 250 ns.  Four different responses under various detunings are plotted in Fig. \ref{fig:simulate1}(a), ranging from $\Delta=5\kappa$ to $\Delta=35\kappa$. The oscillation period follows the detuning, so that a larger detuning yields a higher oscillation frequency, which represents the beating between the laser and WGM frequencies, as discussed in Section \ref{sec:theory}(A). However, the peak height \textit{reduces} when the detuning increases. In Fig. \ref{fig:simulate1}(b) we plot the peak height as a function of  detuning. The data points can be fitted quite well with a Gaussian function and this confirms the presence of the Gaussian term in Eq. \ref{eq:beat}. If the detuning, $\Delta$, is fixed while the ratio $\kappa_e/\kappa$ is changed, according to Eq. \ref{eq:beat}, the peak height should be proportional to this ratio. From the numerical simulations, this linear relationship is confirmed and depicted in Fig. \ref{fig:simulate1}(c). Varying $t_r$ and $\tau$ should not affect the peak height, as shown in both Eq. \ref{eq:beat} and the numerical simulations. However, $t_r$ determines the bandwidth of the transient broadening; therefore, it controls the peak height relationship to detuning, as illustrated in Fig. \ref{fig:simulate1}(d). Significant bandwidth shrinkage is visible only if the rise  time increases by more than a factor of 10. This verifies the behaviour we assumed in Section \ref{sec:theory}. In essence, it is the broadening from the rise time of a detuned pulse that allows light to couple into the WGM and leads to the subsequent beat signal.

\begin{figure}
\resizebox{0.75\textwidth}{!}{
\includegraphics{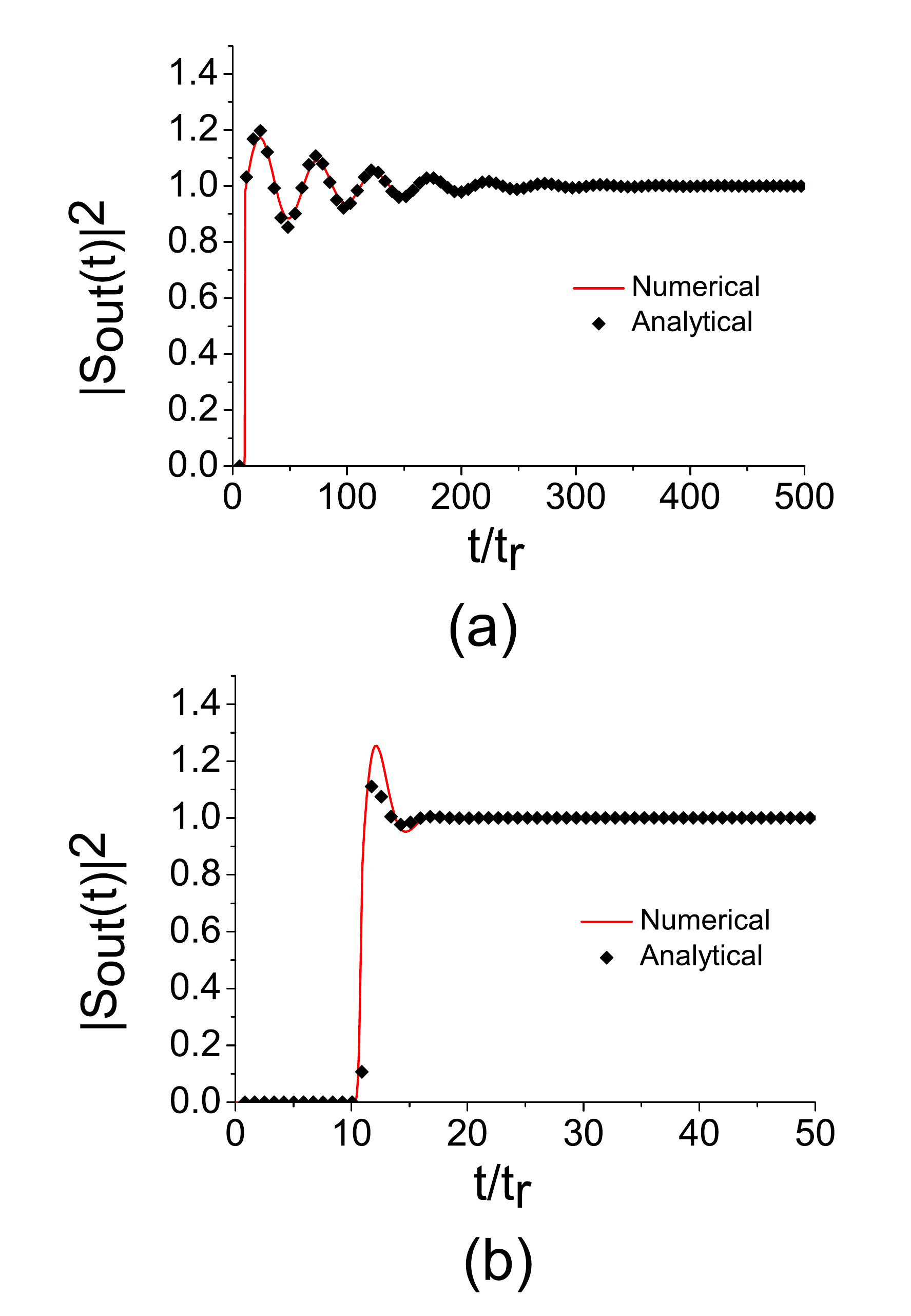}}
\caption{The transient response of a WGR with a pulsed input signal.  Black dots: analytical result using Eq. \ref{eq:beat} and Eq. \ref{eq:beat_at_riseup}.  Red curve: numerical results by solving Eq. \ref{eq:rotateframe} directly. Two different cases were considered. (a) $\tau=200 t_r, \Delta=10\kappa$; (b) $\tau=10t_r, \Delta=2\kappa$.  Both cases satisfy the under coupling condition.}
\label{fig:sim2cal}
\end{figure}

In most cases, we finid that the analytical method gives the same result as the numerical method, see Fig. \ref{fig:sim2cal}(a). Here, the lifetime of the WGM is chosen to be $\tau=100$ ns with a rise  time of $t_r=1$ ns. In this case, $B\gg\kappa$ is well satisfied and both methods yield the same results. However, when the lifetime decreases to $\tau=10$ ns, the analytical result yields a peak height less than the more accurate numerical method (see Fig. \ref{fig:sim2cal}(b)). In this case, $B\sim\kappa$ and the assumption of  narrow filtering by the WGM for obtaining Eq. \ref{eq:beat} is not justified. The broader linewidth of the WGM should permit more photons  to enter into the cavity at the transient broadening time (shown as the shaded area in Fig. \ref{fig:CRUS}(c)), so that, at a later time, more light can beat with the transmitted pulse.

\section{Experiment}
\begin{figure}
\resizebox{\textwidth}{!}{
\includegraphics{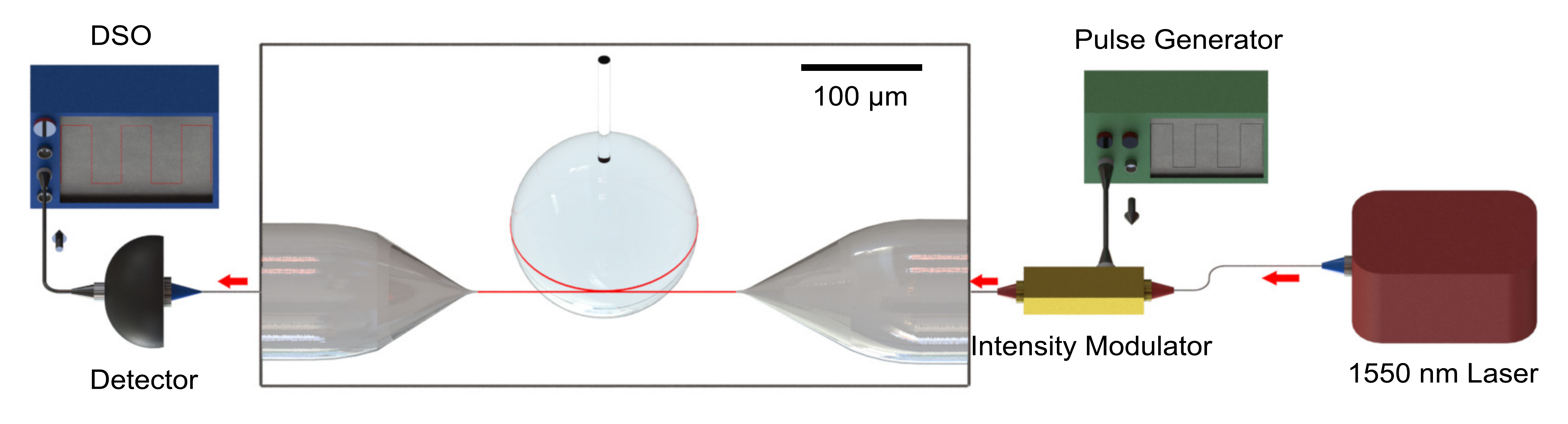}}
\caption{The experimental setup used for cavity ring-up spectroscopy. A 1550 nm laser is intensity modulated and coupled to the microsphere cavity and the transmitted light pules are detected using a fast detector, with the signals recorded on a fast digital storage oscilloscope (DSO).}
\label{fig:RingingExpSetup}
\end{figure}

\begin{figure}
\resizebox{0.75\textwidth}{!}{
\includegraphics{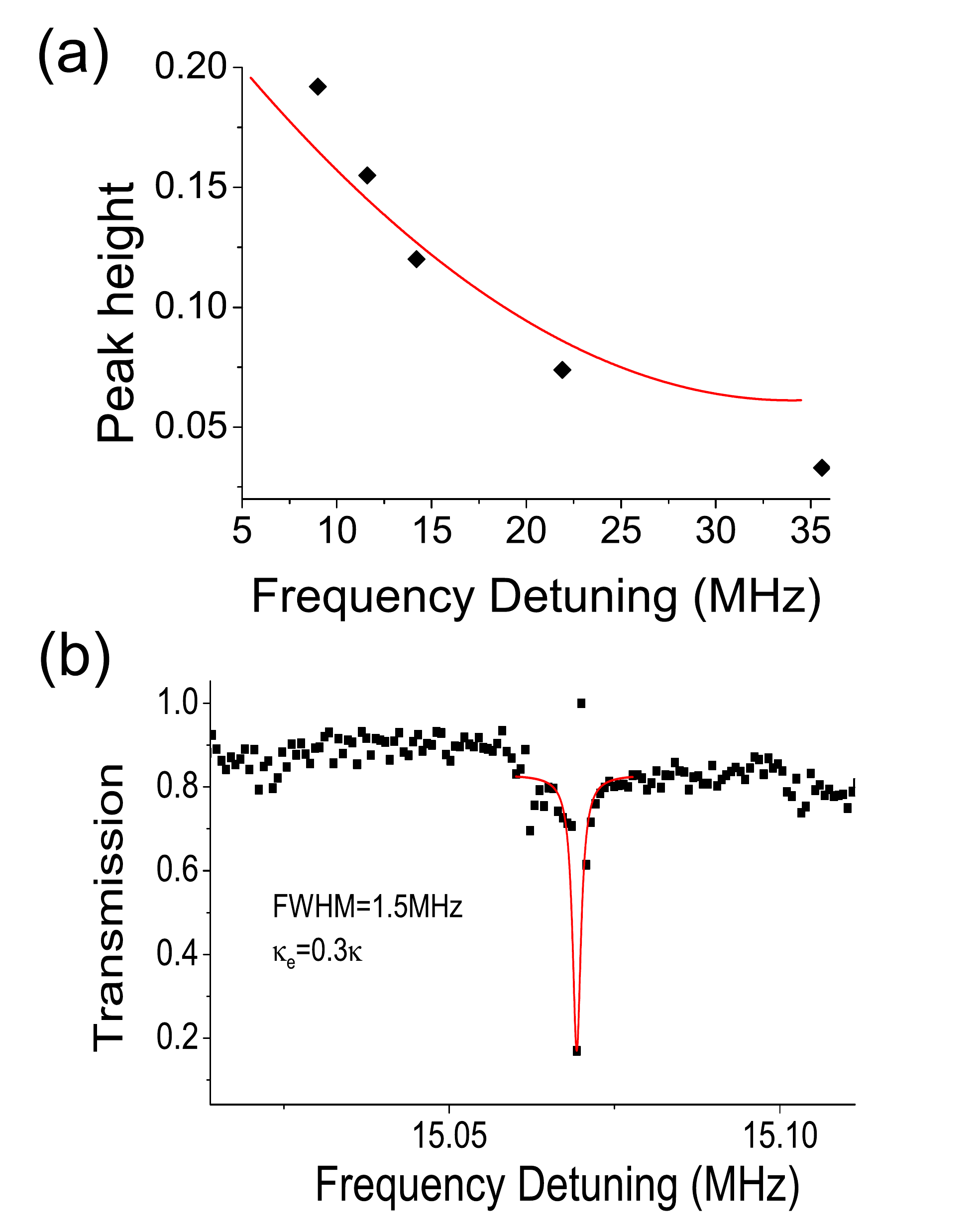}}
\caption{\label{fig:dispersionexp} (a) The peak height measured at different laser detunings simulating a dispersive shift of the microsphere (black dots). The red curve is the theoretical results calculated from the measured parameters. (b) The transmission spectrum of the microsphere which has an ultrahigh \textit{Q} WGM.}
\end{figure}

In order to confirm the above theory, we performed an experiment using an ultrahigh \textit{Q} silica microsphere. The experimental setup is depicted in Fig. \ref{fig:RingingExpSetup}. A 30 $\mu$W, 1550 nm laser was initially modulated using an intensity modulator with an EOM (Thorlabs model LN63S-FC, with rise-up time 50 ps). For this purpose, a pulse generator providing a pulse with a rise time of 5 ns, a width of 500 ns, and a delay about 100 ns was used. The modulated light was coupled to the microsphere's WGM using a fibre taper. The transmission through the fibre was detected with a fast photo detector (Newport model 818-BB-35F) with a typical rise  time of 500 ps. The signal was retrieved on a digital storage oscilloscope (DSO) and recorded at a sampling rate of $>1$ GS/s. The microsphere had a diameter of 80 $\mu$m and the fibre waist was $\sim$1.2 $\mu$m. We chose a high \textit{Q} silica microsphere with a life time of $\sim$500 ns. To study the detuning effects on the CRUS, the microsphere and the fibre taper were aligned to be in contact coupling mode, therefore the coupling condition $\kappa_e/\kappa$ is fixed. The detuning of the laser with respect to the whispering gallery mode was changed so that its frequency approaches that of the WGM in finite steps. The results were normalised to get the peak heights and are plotted in Fig. \ref{fig:dispersionexp}(a).

\begin{figure}
\resizebox{0.75\textwidth}{!}{
\includegraphics{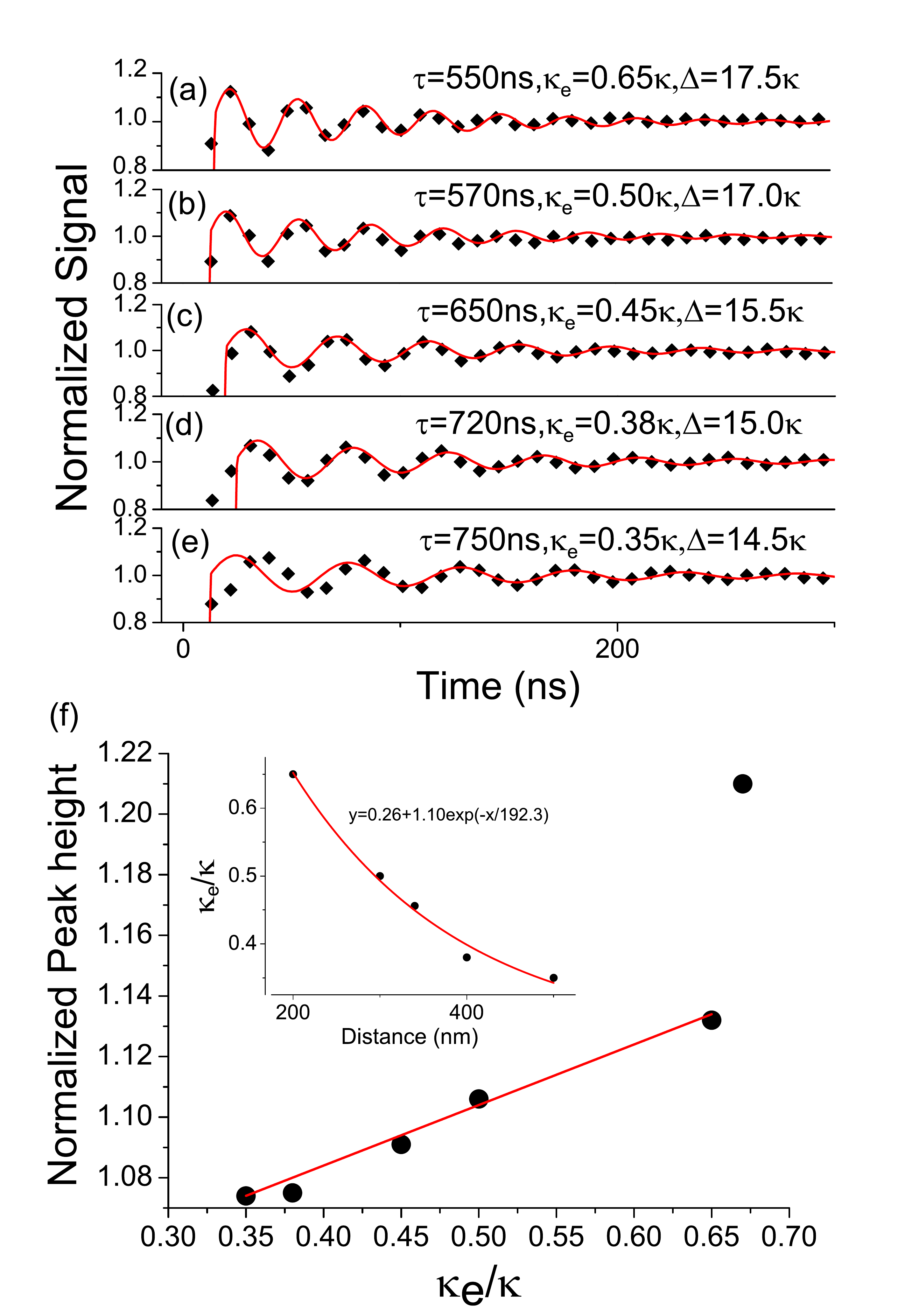}}
\caption{ (a)-(e) The experimental cavity ring-up signals from a silica microsphere resonator. Black dots: Experimental data; Red curves: numerical simulations with a given rise  time, $t_r=5$ ns.  The fitting parameters are given in each case. (f) The peak heights defined as the maxims for plots (a)-(e) satisfy a linear relationship to $\kappa_e/\kappa$. The inset shows the external coupling coefficient as an exponential function of the coupling gap.}
\label{fig:exp}
\end{figure}
To match with the theoretical framework, we made a separate measurement of the transmission spectrum of the WGM being probed, as illustrated in Fig. \ref{fig:dispersionexp}(b). From the transmission efficiency of the mode ($80\%$), and assuming that the system is in the undercoupled regime, it can be deduced that $\kappa_e/\kappa=0.3$. The FWHM of the mode is 1.5 MHz using Lorentz fitting to the dip in Fig. \ref{fig:dispersionexp}(a). In fact, thermal broadening was present even at a low pump laser power of 30 $\mu$W; therefore, $\kappa$ is over estimated. In practice, it is found that $\tau\approx$ 750 ns yields a good fit to the experimental data. By the means mentioned in Sec. \ref{sec:theory}, the theoretical peak height relationship to the frequency detuning is plotted as the red curve in Fig. \ref{fig:dispersionexp}(a). The trend of the peak height to the detuning follows a Gaussian profile.

We also evaluated the peak height with different coupling gaps by varying the gap using a closed-loop, piezo nanopositioner (Smaract SLC1730s-416). The relative position of the taper from the microsphere was determined using a nanopositioner controller (Smaract MCS-3D). The experimental results for different coupling gaps were fitted with the theory to determime $\kappa_e/\kappa$ (see Fig. \ref{fig:exp} (a)-(e)). Since $\kappa_e<\kappa_0$, we assume that $\kappa_e/\kappa=\kappa_e/\kappa_0$. Here, $\kappa_0$ is a constant, while $\kappa_e$ satisfies a near exponential curve to the coupling gap \cite{Gorodetsky1999}, as shown in the inset of Fig. \ref{fig:exp}(f). The corresponding peak height for different coupling conditions plotted in Fig. \ref{fig:exp}(f) shows a near-linear relationship.

From Fig. \ref{fig:exp}(a) to (e), the coupling gap is increased and the period of the CRUS becomes larger. The taper introduces a dispersive red shift to the microsphere's resonance \cite{Madugani2015}. In our experiments, the laser is blue-detuned relative to the resonance and fixed; the larger the distance between the WGR and the taper, the less the dispersion introduced; thus, the cavity mode shifts relative to the laser thereby decreasing the beat frequency. In Eq. \ref{eq:beat}, the peak height should be related to both the coupling condition and the detuning. However, since $\Delta$ appears in the Gaussian term (and assuming that it is large), slightly changing its value will not influence the peak height significantly (see Fig. \ref{fig:simulate1}(b)). In the experiment, we deliberately chose an initial large detuning; therefore, the peak height is still linear with $\kappa_e/\kappa$ despite the dispersive disturbances. When the system is strongly overcoupled, the results deviate from the linear relationship, as seen when $\kappa_e/\kappa=0.75$. In the supplementary material of Ref. \cite{Madugani2015}, it was shown that the dispersive shift rate increases exponentially when moving to a strongly overcoupled regime.  This means that the dispersive influence of the taper will induce a very large  frequency shift for the cavity mode and the changes in the Gaussian term in Eq. \ref{eq:beat} cannot be neglected. As a consequence, the peak height does not vary linearly with $\kappa_e/\kappa$.

\section{Discussion}
Similar to the work in \cite{Shu2015}, the deduced formula of CRUS in this manuscript  shows that it also provides redundant information if the cavity's intrinsic \textit{Q} factor is known; hence, it could be very useful for transient sensing. Instead of doing a time-consuming fast Fourier transform (FFT) of the transient response signal \cite{Rosenblum2015}, one need only record the maximum of the transient signal to retrieve the information for sensing, assuming that one can measure all the other parameters, such as $\kappa_e, \kappa_0$ and $t_r$, from the steady state transmission spectrum. This significantly reduces the complexity of the data processing and decreases the burden for data acquisition. In this sense, the acquisition speed can be further improved. For sensing based on  reactive/dispersive interactions \cite{Vollmer2008}, sensitivity can be optimised by choosing the correct laser detuning. From Fig. \ref{fig:simulate1}(b), the peak height has a Gaussian relationship to the detuning. In order to obtain the highest sensitivity, the pump laser frequency should be chosen so that it is closer to the WGM resonance. For example, in Fig. \ref{fig:simulate1}(b), when $\Delta\sim 10\kappa$ the sensitivity is $dH/d\Delta=0.01\kappa/\Delta$ (where $H$ is the peak height). Also, from Fig. \ref{fig:simulate1}(d), when $t_r$ is longer, the Gaussian profile is steeper; this also improves the sensitivity.

For dissipative sensing \cite{Hu2014},  measuring the peak height for a fixed detuning will also yield valid results. As predicted by theory, the peak height changes  linearly with $\kappa_e/\kappa$. If the system experiences an intrinsic dissipation change due to environmental conditions, the peak height should maintain an inverse relationship to intrinsic dissipation under a certain coupling condition, which is the gap between the taper and the microsphere in our case. In a more complicated scenario, where both dispersion and dissipation exist, a measurement of the peak height may still be sufficient. In the experiment mentioned in the previous section, it was shown that, for large detuning, the peak height always satisfies a linear relationship to $\kappa_e/\kappa$; therefore, it provides a dissipative sensing method immune to any changes to the laser detuning. By having different laser detuning configurations, contributions from dispersive and dissipative interactions can be well categorised.

\section{Conclusion}
In summary, the dynamical mechanisms behind CRUS were investigated by solving the coupled mode equations for a transient response to a Gaussian input pulse. The detailed relationship of the CRUS to laser detuning, coupling coefficient and rise time was determined using approximate analytical solutions. This is further verified by experimental measurements using an ultrahigh \textit{Q} silica microsphere. Using this method, dispersive and dissipative sensing can be performed separately in the transient domain.

\section*{acknowledgments}
This work is supported by the Okinawa Institute of Science and Technology Graduate University. Y. Y and R. M. made equal contributions to the work.

\bibliography{sensing}
\end{document}